\documentclass[fleqn,twoside]{article}
\usepackage{espcrc2}

\usepackage{epsfig}
\usepackage[figuresright]{rotating}

\def\fm{{\rm fm}}
\def\MeV{{\rm MeV}}
\def\GeV{{\rm GeV}}
\def\bs{{\rm B_s}}
\def\fb{F_{\rm B}}
\def\fbs{F_{\rm B_s}}
\def\fps{F_{\rm PS}}
\def\mps{m_{\rm PS}}
\def\cps{C_{\rm PS}}
\def\phirgistat{\Phi_{\rm RGI}^{\rm stat}}
\def\msbar{{\rm \overline{MS\kern-0.05em}\kern0.05em}}
\def\MSbar{{\rm \overline{MS\kern-0.05em}\kern0.05em}}
\def\lamsbar{\Lambda_{\msbar}}
\def\mheavy{M}
\title{
Towards a precision computation of $\fbs$ in quenched QCD
       \thanks{Talk presented by J.~Rolf. Work supported by the 
       EU (contract HPRN-CT-2000-00145) and by the DFG in the SFB/TR
       09. We thank DESY for allocating computer time to this project}
     \thanks{CERN-TH/2003-212, DESY 03-136, HU-EP-03/49, MS-TP-03-8, SFB/CPP-03-20}
   }

\author{J.~Rolf\address[HU]{Institut f{\"u}r Physik, 
    Humboldt-Universit{\"a}t zu Berlin, Newtonstr.~15,
    12489 Berlin, Germany},
  M.~Della Morte\address[DESY]{DESY Zeuthen, Platanenallee 6,
    15738 Zeuthen, Germany},
  S.~D\"urr\addressmark[DESY],
  J.~Heitger\address[MS]{Institut f\"ur Theoretische Physik, 
    Universit\"at M\"unster, Wilhelm-Klemm Str.~9, 48149 M\"unster, Germany},
  A.~J\"uttner\addressmark[HU],
  H.~Molke\addressmark[DESY],
  A.~Shindler\address[DNIC]{NIC/DESY Zeuthen, Platanenallee 6, 15738 Zeuthen, Germany},
  R. Sommer\address[CERN]{CERN-TH, CH-1211 Geneva 23, Switzerland}\,\addressmark[DESY]
  (ALPHA collaboration)
}

\begin{document}

\begin{abstract}
  We present a computation of the decay constant $\fbs$ in quenched
  QCD. Our strategy is to combine new precise data from the static
  approximation with an interpolation of the decay constant around the
  charm quark mass region. This computation is the first step in 
  demonstrating the feasability
  of a strategy for $\fb$ in full QCD. The continuum limits
  in the static theory and at finite mass are taken separately and
  will be further improved.
\end{abstract}

\maketitle

\section{Introduction}

A non-perturbative precise computation of $\fb$ in full lattice QCD
with reliable errorbars would be a major achievement with important
phenomenological implications for CKM physics and the search for new
fundamental processes. Such a computation has to meet three main
obstacles: chiral extrapolation, the heavy b-quark and the large
computational cost due to unquenching. We avoid chiral extrapolations
for the time being by setting the light quark mass to the strange
mass~\cite{mbar:pap3}, thus addressing $\fbs$, which is of interest in
itself, for example in $\bs-\overline{\rm B}_{\rm s}$ mixing. A strategy to
deal with the second problem has been suggested
in~\cite{lat02:rainer,mb:lat01}. The idea is to use HQET and the $1/m$
expansion after a non-perturbative matching to QCD in finite volume.
No large lattices are required and the method can thus be applied to
quenched as well as full QCD. This strategy seems to be viable now
thanks to the considerable improvement of the statistical precision
for computations in the static approximation to QCD that has been
found in~\cite{DellaMorte:2003mn,michele}
The required order of the $1/m$ expansion
is unknown. However, we hope that the size estimates of the terms in
the HQET presented in~\cite{Kronfeldlat03} are correct and the linear
term proves sufficient to keep the systematics under control. This can
with the present computational power only be tested in the quenched
approximation.

Here we compare the renormalization group invariant matrix element
$\phirgistat$ of the static axial current to relativistic data around
the charm quark mass. Later we will also compute the $1/m$ term in the
static approximation to complete the validation of our strategy for
$\fbs$. We use an {\em inter}polation of our relativistic data and of
$\phirgistat$ to obtain a precise quenched value of $\fbs$ in the
continuum limit. Previous results for $\fbs$ have been reviewd
in~\cite{lat02:yamada} and a new computation has recently
appeared~\cite{deDivitiis:2003wy}. 

\section{Numerical Results}

The renormalization group invariant matrix element $\phirgistat$ at
infinite mass is related to the pseudoscalar decay constant $\fps$ at
finite mass by a matching factor $\cps$,
\begin{equation}
  \fps\sqrt{\mps} = \cps(\frac{\mheavy}{\lamsbar}) \times \phirgistat
  + {\rm O}(\frac{1}{m}).
\end{equation}
$\cps$ can be expressed as a function of the renormalization group
invariant heavy quark mass $\mheavy$,
\begin{eqnarray}
  x &=& 1/\log(\frac{\mheavy}{\lamsbar}) \leq 0.62\ \Rightarrow\nonumber\\
  \cps(x) &=&  x^{-\frac{2}{11}} ( 1 - 0.06814\, x - 0.08652\,
  x^2\nonumber\\
  &&  + 0.07939\ x^3).
\end{eqnarray}
This expression is a simple and accurate parametrization of $\cps$
obtained by integration of the entering renormalization group
functions as explained in~\cite{zastat:pap3}, where the anomalous
dimension of the static axial current is taken to 3
loops~\cite{ChetGrozin}. Its uncertainty is estimated to be smaller
than two percent which is half of the difference between the 2-loop
and the 3-loop result.

Our computational setup has been explained in~\cite{fds:JR03,Andreas}.
In particular
we use Schr\"odinger functional boundary conditions and employ 
non-perturbative ${\rm O}(a)$ improvement.

We obtain the decay constant at five meson masses from $\approx
1.7\,\GeV$ to $\approx 2.6\,\GeV$ at four lattice spacings $a$ ranging
from $0.1\,\fm$ to $0.05\,\fm$. Since the data at the different
lattice spacings are not at exactly the same pseudoscalar masses, we
interpolate them linearly in $1/\mps$ and evaluate the interpolating
functions at a few meson masses. These numbers can then be
extrapolated linearly in $a^2$ to the continuum limit. This has to be
done with care however, since the lattice artifacts depend on the
mass. To avoid large discretization errors in the slope of the
continuum extrapolation we thus follow the experience from
perturbation theory~\cite{zastat:pap2} and leave out more and more
coarse lattices at larger masses. At the largest masses only the
finest lattice can be used and we take that point as the continuum
limit. We add to its error the difference to the continuum value
%
\begin{figure}[htbp]
  \vspace{-1cm}
  \centerline{\epsfig{file=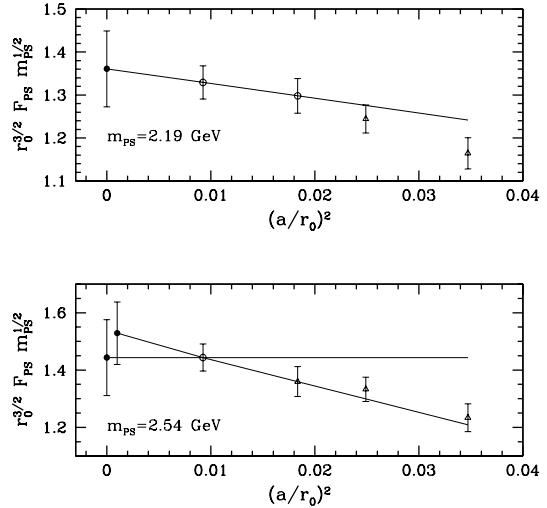,width=\linewidth}}
  \vspace{-1.5cm}
\caption{Continuum extrapolation of the decay constant
  $\fps\sqrt{\mps}$ at two meson masses from two resp. one lattice.
\label{fig:1}} 
\vspace{-0.7cm}
\end{figure}
%
obtained from a linear fit from the two finest lattices. The continuum
limits for two pseudoscalar masses are illustrated in
figure~\ref{fig:1} which shows the case with the two coarsest lattices
left out in the upper part and the case where only one lattice
contributes in the lower part.

Our result is
still preliminary since we will add an even finer lattice with
$a\approx 0.03\,\fm$ (cmp.~\cite{Andreas}) and since the details of
the continuum extrapolations are still being discussed.  Finally we
choose five points close to our original data to be used in the
interpolation between the static and the relativistic case.

To compare with $\phirgistat$ we still have to compute the
renormalization group invariant masses $\mheavy$ that are needed for
the evaluation of the matching factor $\cps$. Here we follow
exactly~\cite{mbar:charm1} for meson masses in the range considered.

We take the static result from~\cite{DellaMorte:2003mn},
\begin{equation}
  \phirgistat = 1.74(13),
\end{equation}
which has been obtained in the continuum limit from three lattice
resolutions with $a\approx 0.1\,\fm \ldots 0.07\,\fm$ with a new
static action that uses HYP smeared links~\cite{HYP}. Furthermore,
wave functions for the states at the Schr\"odinger functional boundary
have been used together with an elaborate technique to reduce the
contribution from the first excited state~\cite{DellaMorte:2003mn} and
the renormalization factor has been computed
non-perturbatively~\cite{zastat:pap3}.

\begin{figure}[t]
  \vspace{-1.cm}
  \centerline{\epsfig{file=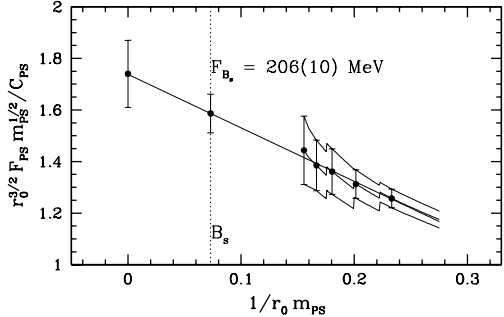,width=\linewidth}}
  \vspace{-1.5cm}
\caption{Comparison of the renormalization group invariant matrix
  element of the static axial current with relativistic data around
  the charm quark mass. \label{fig:2}} 
\vspace{-0.7cm}
\end{figure}

Our results are shown in figure~\ref{fig:2} as a function of the
%
%
inverse pseudoscalar mass. At those points where the number of
lattices taken into account in the continuum extrapolation changes,
there is a small systematic effect which explains the zigzag behaviour of
the 1-sigma band around the relativistic data. If we combine the five
points selected above and the static point by an interpolation in
$1/\mps$ we can obtain $\fbs$. 
To this end we use $M_{\rm b} =
6.96(18)\,\GeV$~\cite{lat02:rainer} and $\lamsbar^{N_{\rm f} = 0} =
238(19)\,\MeV$~\cite{mbar:pap1} to evaluate the matching factor $\cps$
at the b-scale. Both, a linear and a quadratic interpolation lead to
\begin{equation}
  \fbs = 206(10)\,\MeV.
\end{equation}
Here we have used $m_{\bs}=5.4\,\GeV$ and set the scale with
$r_0=0.5\,\fm$.  

Furthermore we notice, that the data are well described by the linear
interpolation that is displayed in the plot.  The slope of the fit is
$-2.1(5)$ while without the constraint through the static
approximation we would obtain $-2.1(1.1)$. To be able to compare this
result with the slope computed from the HQET it is desirable to get
this slope even more precise.

As in~\cite{fds:JR03} we can use the slope of the interpolation to get
an estimate of the quenched scale ambiguity of $\fbs$, i.e. of the
effect of changing $r_0$ within $10\,\%$. Under this change $\fbs$
changes by $12\,\%$. Note that the true quenching error can of course
only be estimated by an unquenched calculation.

\section{Conclusion}

Our comparison of the static and the relativistic results shows that
at the current level of statistical precision nothing 
contradicts the hope that the $1/m$ expansion including only the linear
term
is enough to compute $\fbs$ from the HQET in full QCD at a
sufficient precision. To further validate this strategy we will
compute the $1/m$ term in the quenched approximation from HQET. 

The combination of $\phirgistat$ with our data at finite mass results
in a new value for $\fbs$ in quenched QCD, $\fbs=206(10)\,\MeV$,
including all the systematic errors up to quenching. The quenched scale
ambiguity is estimated by $12\,\%$. Since we {\em inter}polate our
data, this is a stand alone result that is almost independent of any
effective theory, as is indicated by the fact that a linear and a
quadratic function yield identical values.

\bibliography{lattice}
\bibliographystyle{h-elsevier3}

\end{document}